\documentclass[aip,twocolumn,groupedaddress,superscriptaddress,10pt]{revtex4-1}

\usepackage[version=3]{mhchem} 
\usepackage[T1]{fontenc}       
\usepackage{graphicx}
\usepackage{soul,color}

\usepackage{float}

\begin{document}
\title{Resolving and Tuning Carrier Capture Rates at a Single Silicon Atom Gap State}

\author{Mohammad Rashidi}
\thanks{These authors contributed equally to this work.}
\affiliation{Department of Physics, University of Alberta, Edmonton, Alberta, T6G 2J1, Canada}
\affiliation{National Institute for Nanotechnology, National Research Council of Canada, Edmonton, Alberta, T6G 2M9, Canada}
\affiliation{Quantum Silicon, Inc., Edmonton, Alberta, T6G 2M9, Canada}
\email{rashidi@ualberta.net}

\author{Erika Lloyd}
\thanks{These authors contributed equally to this work.}
\affiliation{Department of Physics, University of Alberta, Edmonton, Alberta, T6G 2J1, Canada}

\author{Taleana Huff}
\affiliation{Department of Physics, University of Alberta, Edmonton, Alberta, T6G 2J1, Canada}
\affiliation{Quantum Silicon, Inc., Edmonton, Alberta, T6G 2M9, Canada}
\author{Roshan Achal}
\affiliation{Department of Physics, University of Alberta, Edmonton, Alberta, T6G 2J1, Canada}
\affiliation{Quantum Silicon, Inc., Edmonton, Alberta, T6G 2M9, Canada}
\author{Marco Taucer}
\affiliation{Department of Physics, University of Alberta, Edmonton, Alberta, T6G 2J1, Canada}
\author{Jeremiah Croshaw}
\affiliation{Department of Physics, University of Alberta, Edmonton, Alberta, T6G 2J1, Canada}
\author{Robert A. Wolkow}
\affiliation{Department of Physics, University of Alberta, Edmonton, Alberta, T6G 2J1, Canada}
\affiliation{National Institute for Nanotechnology, National Research Council of Canada, Edmonton, Alberta, T6G 2M9, Canada}
\affiliation{Quantum Silicon, Inc., Edmonton, Alberta, T6G 2M9, Canada}

\begin{abstract}
 We report on tuning the carrier capture events at a single dangling bond (DB) midgap state by varying the substrate temperature, doping type, and doping concentration. All-electronic time-resolved scanning tunneling microscopy (TR-STM) is employed to directly measure the carrier capture rates on the nanosecond time scale. A characteristic negative differential resistance (NDR) feature is evident in the scanning tunneling microscopy (STM) and scanning tunneling spectroscopy (STS) measurements of DBs on both n and p-type doped samples. It is found that a common model accounts for both observations. Atom-specific Kelvin probe force microscopy (KPFM) measurements confirm the energetic position of the DB's charge transition levels, corroborating STS studies. It is shown that under different tip-induced fields the DB can be supplied from two distinct reservoirs: the bulk conduction band and/or the valence band. We measure the filling and emptying rates of the DBs in the energy regime where electrons are supplied by the bulk valence band. By adding point charges in the vicinity of a DB, Coulombic interactions are shown to shift observed STS and NDR features.
\end{abstract}

\maketitle
\section{Introduction}
Whether for the next era of computing, or an improvement of current technology, detailed studies of individual atoms in semiconductors are required~\cite{Morello2010,Koenraad2011}. One important issue is the dynamics of deep levels in the semiconductor band gap~\cite{Pantelides1978,Shockley1952}. In some exotic applications, these deep gap states can act as the device itself. 

STM and charge-sensitive non-contact atomic force microscopy (NC-AFM) are ideal tools to investigate deep gap states on the atomic scale. Owing to the recent development of TR-STM techniques, today it is possible to study the dynamic processes of single atoms and molecules with time resolution down to a fraction of a picosecond~\cite{Nunes1993,Loth2010,Loth2012,Grosse2013,Jelic2017,Moult2011,Saunus2013,Yan2014}. Examples are the measurement of the spin relaxation of individual atoms~\cite{Loth2010}, imaging the ultrafast carrier capture into a single quantum dot\cite{Cocker2013a}, nanosecond resolved study of single arsenic dopants~\cite{Rashidi2016,Rashidia} and femtosecond orbital imaging of a single molecule~\cite{Cocker2016}.
 
One particular technologically relevant gap state is the surface state of a silicon DB. DBs are promising candidates to be employed as building blocks in field controlled computing designs~\cite{Haider2009} and as charge qubits~\cite{Huff2017a,Kolmer2015,Livadaru2010}. They can be placed~\cite{Kolmer2014,Møller2017,Soukiassian.2003}and erased~\cite{AWOACS2017,Pavlicek2017} with atomic precision. Confined quantum-well states can be fabricated by linking DBs in a linear chain~\cite{Schofield2013}. Prior STM and STS analysis have provided both a detailed description of carrier capture by DBs into their gap state~\cite{Berthe2008}, and examination of the Coulomb repulsion energy for a single silicon DB~\cite{Nguyen2010}. Recently, it has been shown that the Coulomb energy separating the two different charge states of a DB, the Hubbard ``U'', gives rise to a characteristic NDR feature~\cite{Rashidi}. In the same study, the carrier capture rate into that midgap state was directly measured by employing all-electronic TR-STM. 

Here, we explore previously inaccessible carrier capture events at a single DB gap-state on hydrogen-terminated Si(100) (H-Si) to thereby expand upon knowledge of and control over silicon bulk to surface state transport. Studies of n and p-type samples reveal that both doping types exhibit a characteristic NDR feature. TR-STM is employed to directly measure the majority carrier transition rate from the bulk to the DB on the nanosecond time scale. The majority carrier concentration is altered by annealing the substrate at different temperatures, with its effect on the capture rate shown. We show that the tip induced band bending can draw the DB's energetic position down to make it resonant with the bulk valence band. In this configuration, the DB can be supplied by the valence band in addition to the conduction band. The filling and emptying rates of the DB in this energy regime are also measured using a variant of all-electronic pump probe methods.  In addition, atom-specific KPFM measurements were used to probe the bias voltages where DB charge state transitions occur.  Furthermore, single point charges were added in the vicinity of a DB to modify both the DB's supply rate from the bulk, as well as the relative energetic position of the charge transition levels of that DB.

\section{Results and discussion}

\begin{figure*}
 \includegraphics[width=12cm]{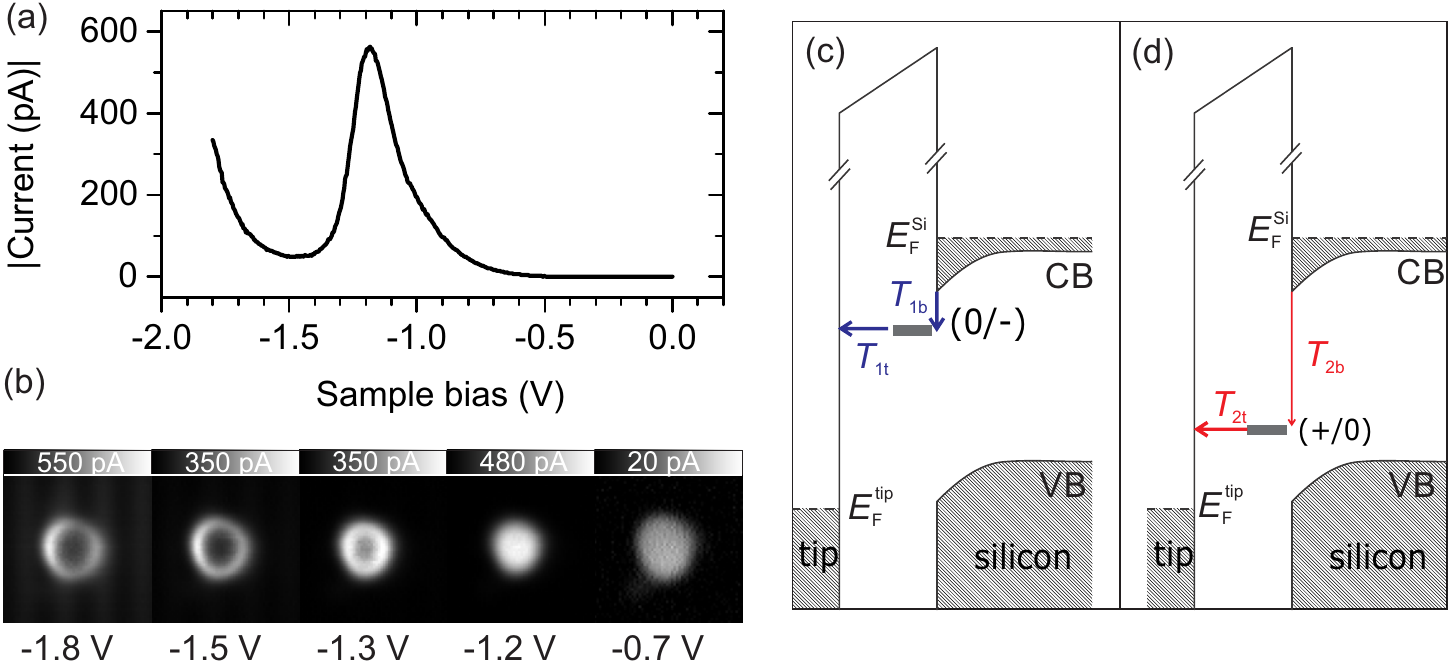}
  \caption{(a) I(V) spectroscopy of a single DB on a degenerately doped n-type H-Si(100) surface showing the characteristic NDR feature at approximately -1.20 V. (b) Constant height STM images of the DB at different bias voltages displaying its characteristic ``ring'' shape in the NDR energy regime. (c) and (d) Energy diagrams of the two conduction pathways in the NDR energy regime. (+/0) and (0/-) denote the charge transition levels of the DB. CB and VB indicate the sample's conduction and valence bands. The Fermi level of the tip and the sample are denoted by $E^{\mathrm{tip}}_{\mathrm{F}}$ and $E^{\mathrm{Si}}_{\mathrm{F}}$. $T_{\mathrm{1b}}$, $T_{\mathrm{2b}}$, $T_{\mathrm{1t}}$ and $T_{\mathrm{2t}}$ denote the time constant of electron transition from the bulk to the (0/-) level, from the bulk to the (+/0) level, from the (0/-) level to the tip and from the (+/0) level to the tip, respectively.}
  \label{fig1}
\end{figure*}

The NDR phenomenon was first observed by Leo Esaki in tunnel diodes~\cite{Esaki1958}.  It has been observed that DBs on degenerately doped silicon surfaces also exhibit NDR~\cite{Berthe2008,Nguyen2010,Rashidi}. Figure 1a displays the I(V) spectroscopy of a DB on a degenerately n-doped hydrogen-terminated Si(100). The tunneling current first rises with increasing negative sample bias voltage, but decreases again (NDR) at approximately -1.20 V. As shown in Fig. 1b, in the NDR energy regime, the DB's center is less conductive than its edge resulting in a ``ring'' shape. A detailed description of the NDR feature and its properties can be found in a previous work~\cite{Rashidi}.  Briefly, when the DB transitions from a predominantly neutral state to a positive condition, a non-resonant inelastic process becomes dominant. This results in a reduced carrier capture rate at the DB.

Figure 1c and 1d display the energy band diagrams of our system in the NDR energy regime. H-Si(100) has no localized surface states. When the surface is degenerately doped on n-type samples, the Fermi level is above the conduction band (impurity band). For this doping condition, DBs are natively negatively charged, i.e., they hold two electrons. Different STM measurement conditions can change the occupation of the DB to hold either one electron (neutral charge state) or zero (positive charge state)~\cite{Labidi2015,Rashidi2016,Taucer2014}.  The charge transition levels are represented by (+/0) and (0/-) in Fig. 1c and 1d. These levels denote the energy required to add the first and second electron to the DB, respectively. 

The energy difference between the charge transition levels arises from the Coulomb repulsion between the two electrons occupying the same DB orbital.  As a result, the two transition levels are mutually exclusive: the DB cannot be singly and doubly occupied at the same time, and the (0/-) transition level is irrelevant unless the DB is already occupied by at least one electron. The  (0/-) level acts as a stepping stone via vibronic coupling~\cite{Berthe2006} for the conduction band electrons to reach the tip, but only as long as the DB holds at least one electron (Fig. 1c). The DB can occasionally be fully emptied by the tip, making it positively charged (Fig. 1d). When the DB is positively charged, the (0/-) level is not available for the electron to tunnel efficiently to the tip. The conduction of electrons from the bulk to the tip stops during this time until one electron transitions from the conduction band to the (+/0) level, making the DB neutral again. Since this process is inelastic, this transition is slow. Direct measurement of the rates shows that the rate electrons pass from the conduction band to the (+/0) level is at least two orders of magnitude slower than the rate to reach the (0/-) level from the conduction band~\cite{Rashidi}. This slow inelastic process is the origin of NDR.

Figure 2a compares I(V) spectroscopy measurements of a DB and H-Si site at decreasing tip-sample distances. Importantly, the DB is conductive in the band gap of silicon. At these energies, the current is through the (0/-) level of the DB, meaning it is always at least singly occupied. Comparing to the spectra acquired over H-Si, we see that there is nearly no measurable current in the gap, except at the closest tip-sample separations. Here, the overlap between the tip and sample wave-functions becomes sufficient to see the current originating from the donor band. These measurements again corroborate that the direct tunneling from the conduction band to the tip is very weak compared to having the DB gap state as a stepping stone.

Figure 2b shows similar curves to Fig. 2a but extended to show the spectroscopy up to a sample bias of -2.0 V. For the tip-height offset of 0 pm, the NDR feature is absent in the I(V) spectroscopy of the DB due to the slow emptying rate from the (+/0) level to the tip~\cite{Rashidi}. As the tip moves closer to the surface (starting with the -100 pm tip height offset), the NDR feature appears at approximately -1.20 V. By increasing the emptying rate with closer tip-sample distances (closer than -300 pm tip-height offset), we are able to make the DB spend a significant fraction of its time in the positive charge state. The positively charged DB becomes clear when the conduction over the DB becomes smaller than H-Si at the same tip height. The point where this occurs is indicated by the arrows in Fig. 2b. The positively charged DB is less conductive than H-Si because the electron conduction from the bulk impurity band to the tip via the DB's (+/0) level is inelastic and therefore slow. At the same time, the charge-induced downward band bending of the valence band in the presence of a positively charged DB reduces the direct tunneling from the valence band to the tip.

\begin{figure}
 \includegraphics[width=6cm]{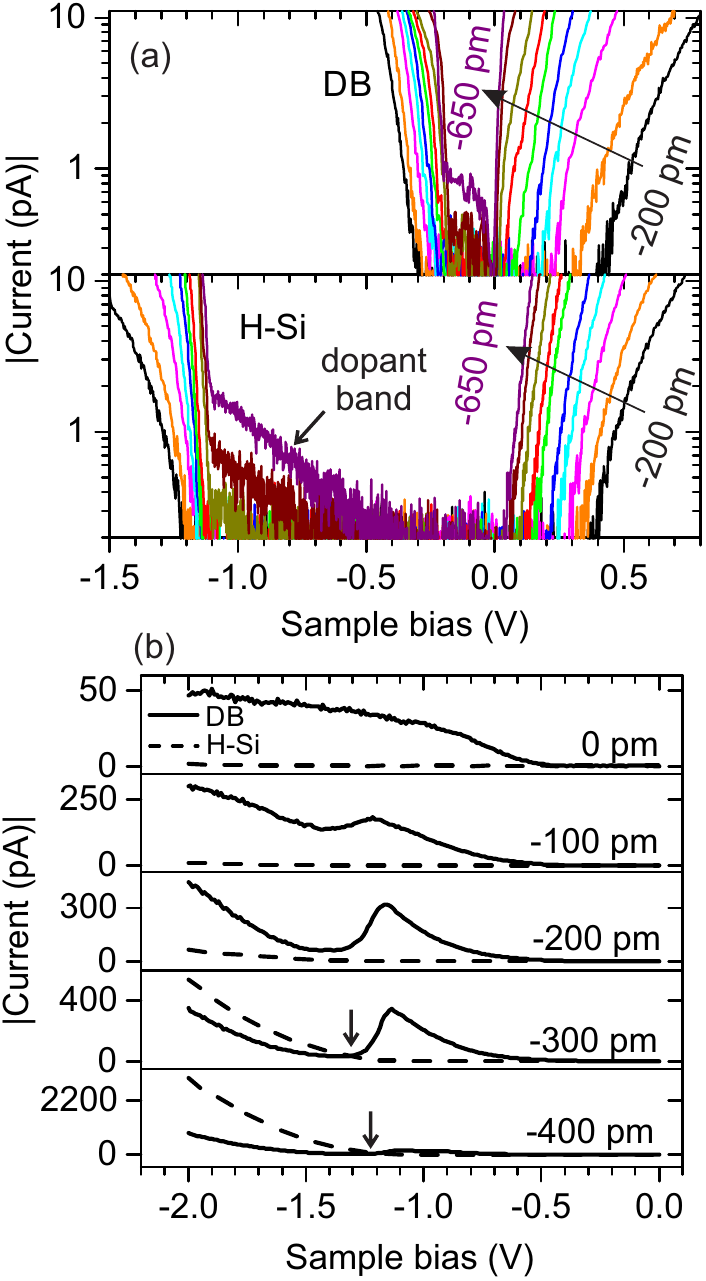}
  \caption{(a) I(V) spectra at DB and H-Si sites with decreasing tip-sample distances.  The initial tip height was set on top of the DB at -1.80 V and 50 pA. The measurements are shown for different tip height offsets between -200 to -650 pm (indicating by black arrows), with 50 pm intervals. More negative tip height indicates a smaller tip-sample separation. The current originating from the dopant (impurity band) is labeled in the bottom panel. (b) I(V) spectra on a DB (different from the DB presented in (a)) with a wider bias window. The initial reference tip height (0 pm) was set on top of the DB at -2.00 V and 50 pA. Tip height offsets are shown to the right of each curve. The arrows indicate the crossing point where conductivity at the DB becomes less than at H-Si.}
  \label{fig2}
\end{figure}

The KPFM measurement in Fig. 3a shows the bias voltages where the charge transitions on an n-type sample occur. The step-like shift in the AFM frequency response at approximately -0.20 V corresponds to a single electron charge transition. The DB goes from a negative charge state left of the step, to a neutral charge state on the right~\cite{Huff2017a}. We also observe a secondary feature just before the onset of the NDR regime, a small dip in the frequency response at approximately  -1.10 V. This dip indicates that, for this time averaged measurement, the charge is temporarily positive---consistent with our I(V) results. The features observed in KPFM measurements occur at lower biases than the corresponding current shifts in the I(V) spectroscopy. This is because KPFM is sensitive to single electron charge changes, whereas I(V) spectroscopy measures the electronic rates. It is possible that the tip Fermi level is already past the charge transition levels of the DB, but because of small electronic rate changes I(V) does not show an immediate response. We expect that the KPFM measurements are much better suited to show the relative energetic positions of the DB's charge transition levels at a given tip height. This is consistent with previous studies where the capability of KPFM to resolve single electron charge transitions in charged metallic species~\cite{Konig2009,Steurer2015b}, single electron transfer between molecules~\cite{Steurer2015a}, and charge state transitions in quantum dots were shown~\cite{Stomp2005}.

\begin{figure}
 \includegraphics[width=6cm]{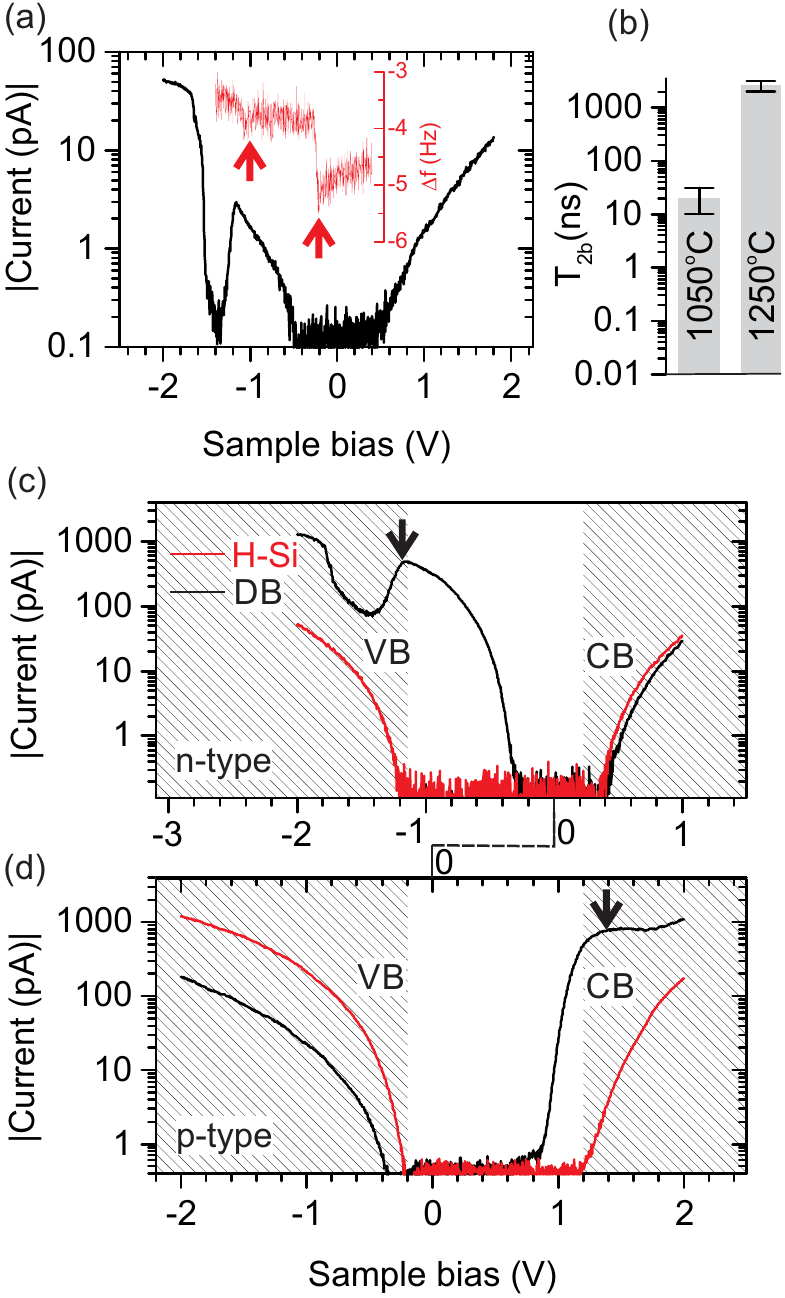}
  \caption{(a) KPFM (red) and NDR I(V) (black) curves measured for a DB on a 1250$^o$C annealed sample. The KPFM curve was measured at z=-330 pm from the reference height of -1.80 V and 50 pA, set over hydrogen. The oscillation amplitude was 100 pm. Two features in the KPFM curve are visible (indicated by red arrows) which correspond to the (0/-) charge transition level (approximately -0.20 V) and the (+/0) charge transition level (approximately -1.10 V).  (b)  $T_{2\mathrm{b}}$ time constant (see Fig. 1d) of DBs on samples annealed at different temperatures (1250$^o$C and 1050$^o$C) during preparation. (c) and (d) show the symmetric properties of the rate-limited bulk-to-surface transport effect in n-type and p-type silicon, respectively.  The I(V) curves comparing a DB with H-Si show the characteristic feature of NDR when the charge carriers are electrons (n-type) and also holes (p-type).  Note, the graphs have been shifted to align the bulk conduction and valence band edges of both curves. The bands are hatched for clarity and the NDR onset features are marked with a black arrow for both n and p-type.}
  \label{fig3}
\end{figure}

We are able to control the carrier density at the surface region by changing the bulk doping concentration through our choice of sample annealing temperature, or through measurement at different ambient temperatures. Implementing all-electronic time resolved techniques~\cite{Rashidi}, we show the effect on the electron capture time constant from the bulk, shown schematically in Fig. 1b. The greater the density of carriers, the faster these rates will be. This is because the carrier capture rate from a mid-gap state ($\Gamma$) is equal to $n\gamma$, where $n$ is the majority carrier density in the bulk conduction or valence band, and the coefficient $\gamma$ is related to the carrier thermal velocity and capture cross-section of the mid-gap state~\cite{DelaBroise2000}. We control the concentration of dopants at the surface of a material by annealing the sample to different temperatures during preparation~\cite{Labidi2015,Pitters2012,Rashidi2016}. Higher annealing temperatures cause a depletion of dopants in the near surface region on n-type and thereby decrease the available carriers at the surface. Figure 3b compares the electron capture rates of 1050$^o$C and 1250$^o$C annealed samples. As expected, DBs on higher annealed samples have slower electron capture by two orders of magnitude. This is consistent with the secondary ion mass spectroscopy measurements presented in previous studies~\cite{Labidi2015,Pitters2012}. We can also thermally induce carriers by increasing the temperature of the sample during measurements. At room temperature and 77 K, there are sufficient carriers for the electron capture rate to become too fast for our experimental setup to resolve. 

In Fig. 3c and 3d we observe the complementary nature of the properties of n-type and p-type semiconductors. I(V) experiments of DBs on p-type samples exhibit a NDR feature at a positive sample bias close to the conduction band edge. To emphasize the similarity of mechanism, the bias voltages in Fig. 3a and 3c are shifted to make the bulk conduction and valence band edges aligned in both plots. The NDR on p-type samples can be explained by the same mechanism as the n-type samples, but by replacing: but by replacing electron capture from the bulk conduction band with hole capture from the valence band. In addition, electron injection from the tip to the DB level can be viewed as hole emptying by the tip. All other arguments about the tip height and charge transition levels are the same as a n-type sample. We note that the same NDR on DBs on p-type samples has been observed before on boron $\delta$-doped Si(111)~\cite{Berthe2008,Nguyen2010}. We were not able to resolve the hole capture rate of DBs on the p-type samples, possibly because it was much faster than the time resolution of our set up (<1 ns).

\begin{figure*}
 \includegraphics[width=12cm]{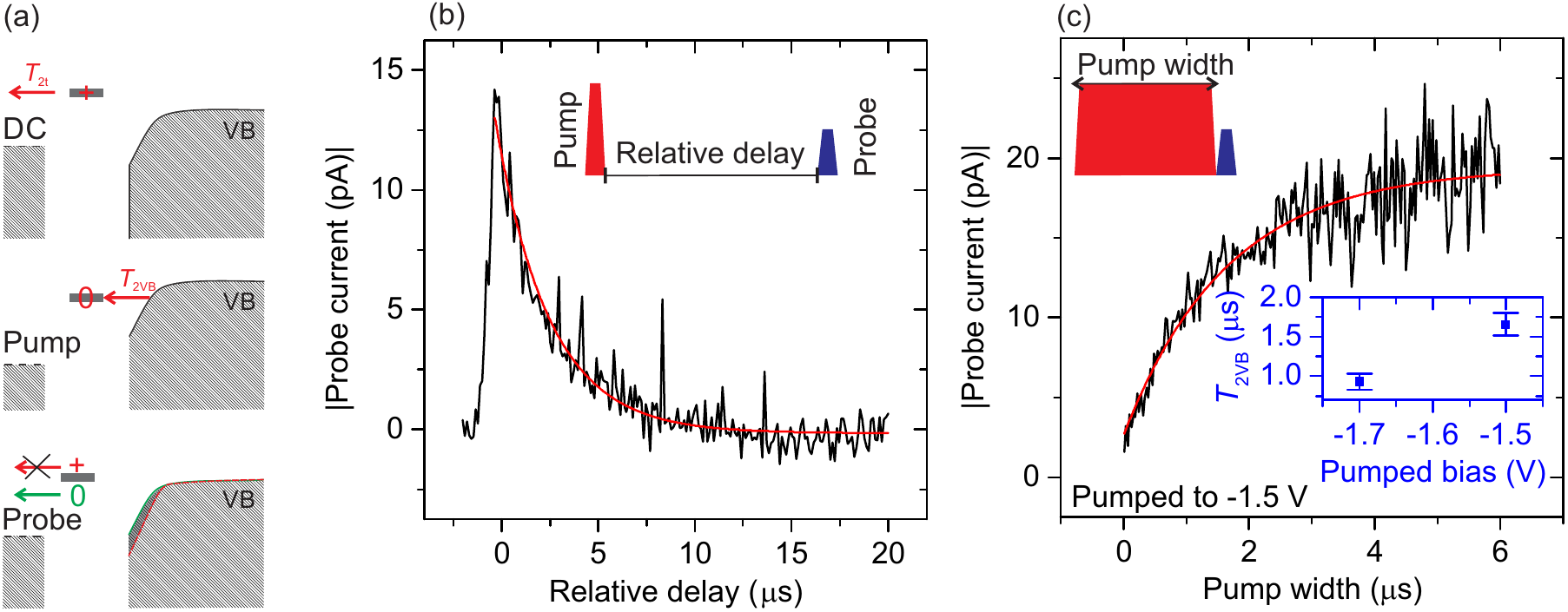}
  \caption{(a) Schematic showing the measurement of the filling and emptying rates of the (0/+) level from the valence band to the tip using TR-STM.  DC represents the fixed bias used to empty the DB to a positive state.  Pump represents the pulse bias used to fill the DB from the valence band, and probe represents the pulse bias used to sample the charge state of the DB.  $T_{\mathrm{2VB}}$ denotes the time constant of electron transition from the valence band to the (0/+) state.  $T_{\mathrm{2t}}$ denotes the time constant of electron transition from the (0/+) state to the tip.  (b) Measurement of $T_{\mathrm{2t}}$ by sweeping the relative delay between the pump and probe pulse. The red curve is the single exponential fit. The inset schematically illustrates the measurement setup. (c) Measurement of $T_{\mathrm{2VB}}$ by varying the pump pulse width, while maintaining the probe pulse height and relative delay constant. The blue inset is the measurement of $T_{\mathrm{2VB}}$ for different pumped biases. The red curve is the single exponential fit. The top inset illustrates the measurement setup. For the experiments shown in (b) and (c), a 1250$^o$C flashed sample is used.}
  \label{fig4}
\end{figure*}

Past the NDR regime, at higher negative bias, there is a sharp turn on in current corresponding to the (+/0) level becoming resonant with the valence band~\cite{Rashidi}. As shown in Fig. 4a, at this energy, a new filling rate ($1/T_\mathrm{2VB}$)  from the bulk is introduced. We use an all-electronic pump probe technique to measure this rate. The schematic in Fig. 4a outlines the three elements of the experiment: the DC bias, the pump, and the probe. The DC bias is set at a value in the NDR region, and thus works to empty the DB. The pump pulse height is set to a value that pulls the (+/0) level into resonance with the valence band, serving to fill the DB. Finally, the probe is set at an energy near the valence band threshold to probe the state of the system. If the level is empty, the corresponding downward band bending from the positive charge means the probe will collect less current. If the level is neutral or full, then the VB edge at the surface will be higher, and the probe will collect more current. 

The emptying rate to the tip can be measured by sweeping the relative delay between pump and probe (schematic in Fig. 4b). By staying at the DC bias for longer amounts of time we measure an exponential decrease in current. The time constant associated with this decay is the emptying rate, $T_\mathrm{2t}$, at the valence band edge. The pump pulse must be long enough to ensure that the (+/0) level is occupied with an electron. For the data shown in Fig. 3b we set the pump pulse width to 300 ns, the DC Bias to -1.2 V, the pump amplitude to -0.4 V, the probe pulse to 1 $\mu$s and the period of the pulse trains to 50 $\mu$s. The time constant we measured for the emptying rate at this energy is $T_{2\mathrm{t}}$=2.78$\pm$0.12 $\mu$s consistent with our previous measurements on a 1250$^o$C flashed sample. 

To measure the filling time constant ($T_{\mathrm{2VB}}$), we keep all parameters constant except for the pump width. For small pump widths the (+/0) level has not had the chance to be filled by valence band electrons, resulting in the probe pulse measuring less current. For larger pump widths the (+/0) level has sufficient time to be filled by the valence band, resulting in more measured current by the probe pulse. By sweeping this value and measuring probe current, we get the exponential curve shown in Fig. 4c. Note that since we are changing the duration of the pump pulse, we must be careful in extracting the probe current from the averaged measured current. To do this we run two experiments. One with both the pump and probe, and one with just the pump. We then take the difference in the collected current for these measurements to extract only the probe current. The data shown in Fig. 4c is the probe current only. By fitting this curve we extract the filling time constant ($T_{\mathrm{2VB}}$). The relative delay between pump and probe pulses must be constant and smaller than the emptying time of the tip. For the experimental data shown in Fig. 4c we set the relative delay to 10 ns, the DC bias to -1.3 V, the probe amplitude to -0.18 V, and the period of pulse trains to 50 $\mu$s. The time constants measured for different pumped voltages are shown in the blue inset of Fig. 4c. As expected, the filling rate increases with higher pump amplitude because the overlap of the (+/0) level with the valence band increases. For the experiments shown in Fig. 4, a 1250$^o$C flashed sample was used. 

\begin{figure}
 \includegraphics[width=7.5cm]{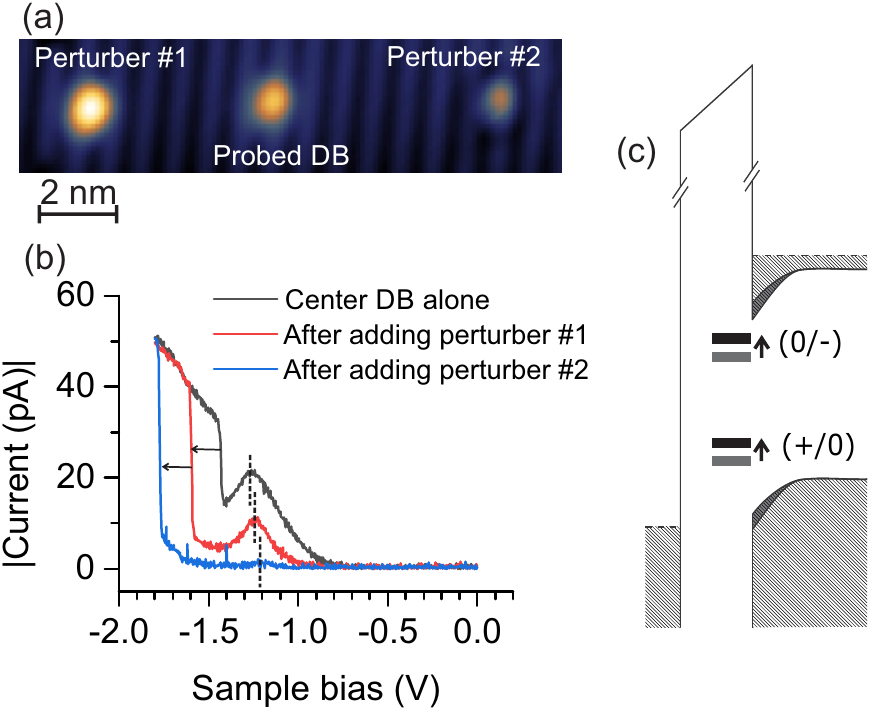}
  \caption{(a) STM image of the initial central DB and the two perturbing DBs.  (b)  The I(V) spectra of the lone central DB before (black), after the addition of perturber \#1 (red), and after the addition of perturber \#2 (blue). The dashed lines indicate the onset of NDR for each curve. The arrows indicate the shift of the post-NDR current onset after adding each perturber DB. (c) Energy band diagram of the system of study displaying the energy shift of (+/0), (0/-), the conduction band and the valence band in the presence of a perturbing DB. }
  \label{fig5}
\end{figure}

The transition levels are defined as the energy required to put additional electrons into the DB. This implies that the effective electric field at the site will energetically shift the states. The presence of any negative charge, such as another DB, will cause the transition levels and the surrounding bands to be shifted higher in energy as shown in Fig. 5c. This is intuitive since it will take more energy to add an electron to the DB if there are other negative charges in the vicinity.  By adding DBs around a target DB, as shown in Fig. 5a, we observe a consistent shift of features in the I(V) curves. The black curve in Fig. 5b shows the spectroscopy of the center DB in the image of Fig. 5a before any additional DBs were added. The two surrounding DBs were sequentially placed in the order they are labeled with a corresponding I(V) taken over the probed DB after each addition.

We can understand these shifts by discussing each of its features and their relation to the newly positioned transition levels. First, the amplitude of the NDR peak decreases indicating that the conduction band is being bent upwards. There are now fewer states below the Fermi level to supply the DB. The second feature is the onset of the NDR region, which shifts to a less negative bias as indicated by the vertical dashed lines marking the onset of the NDR in Fig. 5b.  This is due to the (+/0) level now being higher in energy, so that it can be emptied by the tip at smaller negative sample biases. In the NDR region the current tends to 0, and the DB is on average empty because the filling rates from the bulk have decreased. Finally, the post-NDR onset of the valence band current shifts to more negative bias values since the (+/0) level now requires a larger negative bias to come into resonance with the valence band.

\section{Conclusions}
To summarize, we used the techniques of STS, KPFM, and TR-STM to examine the effect of substrate temperature, doping type, doping concentration, and electrostatic perturbation on a single DB gap state. The NDR effect originally observed with n-type samples was found to also result with p-type samples under complimentary conditions and that could be straightforwardly accounted for by extension of a common model. STS measurements illustrate that controlled change of the tip-sample distance allows control of the average charge state of the DB and conduction through that state.  KPFM corroborates the STS study, giving the relative energetic positions of the DB's charge transition levels. Further tuning was achieved by altering the majority carrier concentration of the system. Measurements made at higher temperature enabled thermionic carrier generation and the increase of rates. The first measurement of filling and emptying rates for the (0/+) level from the valence band have been determined. Finally, a target DB's electronic character was shown to be tunable in a predictable manner through Coulombic interactions induced by the placement of negatively charged DBs nearby.

\section{Methods}

These measurements were performed using an Omicron Low temperature STM operated at 4.5 K, 77 K and room temperature. A Nanonis SPM controller and associated software were used for data acquisition. STM tips were made of both tungsten and iridium to demonstrate our results are tip independent. Tungsten and iridium were electrochemically etched, cleaned, and sharpened by nitrogen-assisted field ion microscopy\cite{Rezeq2006}. Boron (5-7 m$\Omega$.cm) and arsenic (3-4m$\Omega$.cm) doped Si(100) wafers were used in these experiments. The samples were degassed for several hours at ~600$^o$C prior to hydrogen termination.  The oxide layer was then desorbed by flash annealing the crystal between 1050$^o$C and 1250$^o$C, depending on desired dopant concentration. Hydrogen termination was done at 330$^o$C under the exposure of H atoms for 30 seconds. Single DBs are created by placing the tip over a hydrogen atom at a reference height defined by 1.30 V and 50 pA, and applying a positive sample voltage pulse (2.0 to 2.4 V) to desorb the hydrogen atom.

Radio frequency (RF) wiring with a 500 MHz bandwidth enables the STM to achieve all-electronic time resolved measurements on the order of nanoseconds. A RF switch (Mini-Circuits ZX80-DR230-S+) was connected to each of the two output channels of an arbitrary function generator (Tektronix AFG3252C), toggling between it and ground. The outputs of the switches were fed into an adder (Mini-Circuits ZFRSC-42-S+) and connected to the tip. To account for distortion that may occur for small pulses at the junctions, cross-correlation signals were measured over H-Si to extract a proper calibration. To mitigate ringing at the junction, the pulse edges were set to 2.5 ns.  In our time-resolved experiments we can vary many parameters including: pulse frequency, pulse amplitude, pulse width, and relative delay between pulse pairs from each output.

NC-AFM experiments were all carried out at 4.5 K. A Tungsten-tipped commercial (Omicron) qPlus AFM sensor with a separate tunneling wire was used to avoid cross-talk problems under bias~\cite{Majzik2012}. The sensor exhibited a quality factor of 15k with a resonance frequency of 25 kHz. KPFM curves were taken from a fixed height with a fixed amplitude of 100 pm, calibrated using the tunnel current method~\cite{Simon2007}. The tip was left to settle overnight before all curves were acquired to minimize piezo drift.

\begin{acknowledgements}
We thank Martin Cloutier and Mark Salomons for their technical expertise. We also thank NRC, NSERC, AITF, CRC, CIFAR, and Compute Canada for support.
\end{acknowledgements}

\end{document}